\documentclass[12pt]{article}

\usepackage{amsmath,amssymb}
\usepackage{cite}

\begin{document}

\title{The approach to gravity as a theory of embedded surface}

\author{
A.A.~Sheykin\thanks{E-mail: anton.shejkin@gmail.com},
S.A.~Paston\thanks{E-mail: paston@pobox.spbu.ru}\\
{\it Saint Petersburg State University, Saint Petersburg, Russia}
}
\date{\vskip 15mm}
\maketitle

\begin{abstract}
We study the approach to gravity in which our curved spacetime is considered as a surface in a  flat ambient space of higher
dimension (the embedding theory). The dynamical variable in this theory is not a metric but an embedding function. The
Euler-Lagrange equations for this theory (Regge-Teitelboim equations) are more general than the Einstein equations, and
admit "extra solutions" which do not correspond to any Einsteinian metric. The Regge-Teitelboim equations can be
explicitly analyzed for the solutions with high symmetry. We show that symmetric embeddings of a static spherically symmetric asymptotically flat metrics in a 6-dimensional ambient space do not admit extra solutions of the vacuum Regge-Teitelboim equations. Therefore in the embedding theory the solutions with such properties correspond to the exterior Schwarzchild metric.
\end{abstract}

\newpage

\section{Introduction}
The properties of surfaces embedded in an ambient space are studied since the 19th century. After the introduction of the special relativity it has become possible to reduce the study of dynamics to the study of some surface in spacetime. For a point particle it is an one-dimensional curve -- a worldline, for a string -- a world sheet, etc. The ADM formalism for gravity constructed in the early 60s \cite{adm} can be also considered as a dynamics of a 3-dimensional surface embedded in a (3+1)-dimensional spacetime.

The next step was done by Regge and Teitelboim. By analogy with the string theory, which is a theory of a 2-dimensional surface in an  ambient spacetime, they proposed \cite{regge} to consider our 4-dimensional curved spacetime as a surface in the Minkowsky spacetime of higher dimension.
The reasons for such a consideration are mainly the difficulties which arise in the canonical quantization of the General Relativity. These difficulties include the invariance of the theory under the time reparametrization, which reduces the Hamiltonian to the combination of constraints; the impossibility of the causality determination related to the operator nature of the metric, etc. Note however  that Pav$\check{\text{s}}$i$\check{\text{c}}$, independently of Regge and Teitelboim, developed a slightly different theory \cite{pavsic85,pavsic85let} in order to  eliminate conceptual rather than technical problems. In his opinion, such a theory could solve the problem of incompatibility of the quantum mechanics with the existence of an "objective reality" in the spacetime. A similar theory was also independently constructed in \cite{estabrook1999}. Following the authors, the main advantage of this theory is a simplification of constraints. In \cite{statja25} the embedding theory is formulated as a field theory in the Minkowsky spacetime.

Whatever the motivation, technically the consideration of the gravity as an embedding theory is equivalent to the change of variables in the action, which makes the embedding function $y^a(x^{\mu})$ a dynamical variable, instead of  the metric $g_{\mu\nu}(x^{\mu})$:
\begin{align}
      g_{\mu\nu}(x) = \partial_\mu y^a(x) \partial_\nu y^b(x) \eta_{ab},\label{met}
\end{align}
so $g_{\mu\nu}(x)$ becomes induced. Here $\mu,\nu = 0 \ldots 3$, $\eta_{ab}$ is the  metric of the $N$-dimensional flat ambient spacetime, $a,b = 0 \ldots N-1$.

According to the Friedman theorem \cite{fridman61}, the minimal ambient spacetime dimension required for a local isometrical embedding of a  $n$-dimensional curved spacetime, in the general case is equal to $n(n+1)/2$. Thus the equivalence with the General Relativity can be achieved if $N\geq 10$. For that reason theories with $N=10$ are mainly considered, although, for example, a formalism with $N=14$ was also developed \cite{maia89,bustamante, willison}.

It should be noted here that initially the embedding theory was not considered as a new physical theory, but rather as an alternative formulation of the General Relativity, which potentially might ease the quantization. This is a fundamental distinction with the brane theory \cite{lrr-2010-5} which appeared slightly later. Although both theories consider our spacetime embedded in the bulk, in the embedding theory the bulk is necessarily free of the gravity, otherwise we would be forced to deal with all the difficulties mentioned above. On the contrary, the brane theory was originally constructed as an extension of the Standard Model, designed to solve some problems of the particle physics, without direct relation to the quantization of the gravity itself. However, there were  some attempts to unify the Brane-World and the Regge-Teitelboim (RT) approach, see, for example, \cite{davgur06}.

\section{The Regge-Teitelboim equations and\\ their extra solutions}
Since  the main purpose of the RT approach is not a search for a new physical theory, but a reformulation of an already existing one, the action in this approach can be chosen as the Einstein-Hilbert one (for some arguments about the addition of higher curvature terms see, e.g., \cite{kokarev1998}). One can make a substitution (\ref{met}) in the variation of the action and integrate it by parts:
\begin{multline}
      \delta S = \frac{1}{2\varkappa} \int d^4 x  \sqrt{-g} (G^{\mu\nu} - \varkappa T^{\mu\nu}) \delta g_{\mu\nu}
       = \frac{1}{\varkappa} \int d^4 x  \sqrt{-g} (G^{\mu\nu} - \varkappa T^{\mu\nu}) \eta_{ab} \partial_\mu y^a \partial_\nu \delta y^b  = \\ = -\frac{1}{\varkappa} \int \partial_{\mu}( \sqrt{-g} (G^{\mu\nu} -\varkappa T^{\mu\nu} ) \partial_{\nu} y^a) \delta y_a.     \label{var_RT}
\end{multline}
Equating this variation to zero, one can obtain the Regge-Teitelboim equations \cite{regge}:
\begin{align}
      \partial_{\mu}( \sqrt{-g} (G^{\mu\nu} -\varkappa T^{\mu\nu} ) \partial_{\nu} y^a) = 0,\label{RT}
\end{align}
where the Einstein tensor $G^{\mu\nu}$ and the energy-momentum tensor $T^{\mu\nu}$ now depend on the embedding function $y^a(x)$.

These equations contain the derivative of the Einstein tensor, which itself contains the second derivatives of the metric, and the metric is expressed through the derivatives of the embedding function. A naive power counting gives a reason to suppose that the RT equations contain the fourth order time derivatives, which are physically undesirable. But actually the order of time derivatives in these equations is no higher than two \cite{deser}.  In order to prove  that one can transform the derivative in (\ref{RT}) into a covariant one and apply the Bianchi identities $D_\mu G^{\mu\nu}=0$ and the covariant conservation law $D_\mu T^{\mu\nu}=0$. After that the equation (\ref{RT}) takes the form
\begin{align}
 (G^{\mu\nu} -\varkappa\, T^{\mu\nu} ) b^a_{\mu\nu} = 0,\label{RT1}
\end{align}
where $b^a_{\mu\nu} \equiv D_{\mu} \partial_{\nu} y^a$ is the second fundamental form of the surface, which contains only second derivatives of the embedding function. Remembering that the Einstein tensor, as the curvature tensor, can be algebraically expressed through $b^a_{\mu\nu}$, one can conclude that the RT equations in the form (\ref{RT1}) contain time derivatives of no higher than second order.

However, these equations are obviously more general than the Einstein equations,  there are the so-called "extra solutions" for which $G^{\mu\nu} \neq \varkappa\, T^{\mu\nu}$. The reason of this lies in the restriction of the class of functions which are varied in (\ref{var_RT}).
The existence of a derivative in the substitution (\ref{met}) leads to the appearance of extra differentiation in the field equations.

These extra solutions can be treated as a disadvantage which needs to be eliminated. For this purpose, e.g., in \cite{bustamante,willison} there was proposed the formalism of the so-called "free embeddings" in the $14$-dimensional spacetime. In that case the $b^a_{\mu\nu}$ behaves in a manner of a square matrix: it contains 10 independent components by index $a$ (four out of 14 vanish due to the transversality condition $b^{a}_{\mu\nu} \partial_{\alpha} y_a = 0$), while the  two lower indices, by which it is symmetric, can be considered as one multi-index which runs over 10 values. If the determinant of such $10\times 10$ matrix is not  equal to zero, then  the equations (\ref{RT1}) become equivalent to the Einstein equations.

Among the attempts of eliminating the extra solutions one should note the papers \cite{faddeev}, in which the dynamical variable is not the embedding function but its derivative, so the field equations do not  contain the extra derivative. Thus the extra solutions of the field equations are absent, but a torsion is appeared, and its behavior needs to be examined.

The most straightforward way of removing the extra solutions is the imposing the so-called "Einsteinian constraints"
\begin{align}
G_{\mu\bot}-\varkappa\, T_{\mu\bot}=0     
\end{align}
where the symbol $\bot$ denotes the direction orthogonal to the constant time surfaces. This possibility was mentioned in the original Regge and Teitelboim paper \cite{regge}. In \cite{statja18,statja24} the authors have  shown that imposing these constraints at the initial time is sufficient in order to  achieve the equivalence with the General Relativity. In \cite{statja18} was also proposed a possible action of the embedding theory with additional constraints, but this action looks quite artificial and is not  Lorentz-invariant.

The fact that the existence of the extra solutions depend on the initial values can be easily understood if one  consider the following toy model. Substituting  $q(t)=\dot{y}(t)$ into the action for a harmonic oscillator
\begin{align}
      S=\int dt \left(\frac{\dot{q}^2}{2}-\frac{\omega^2{q}^2}{2}\right) \label{osc1}
\end{align}
and considering  $y(t)$ as a new dynamical variable, we vary the action (\ref{osc1}) with respect to $y(t)$ and we obtain the field equation
\begin{align}
\frac{d}{dt} \left(\frac{d^3y}{dt^3}+\omega^2 \frac{dy}{dt} \right)=0. \label{osc2}
\end{align}
Integrating this equation and returning to the old variable $q$, one finds that
\begin{align}
\ddot{q}+\omega^2 q =C, \label{osc3}
\end{align}
where $C$ is a constant.
For the full equivalence of this theory and the harmonic oscillator it is  enough to require the satisfaction of the oscillator equation of motion at some time, then $C=0$ permanently. In a field theory the extra solutions are governed not by a single constant, but by a set of some conserved currents, though the general situation remains the same -- if these currents vanish at some moment, the conservation laws ensure their vanishing at any time.

Despite the fact that from this point of view the extra solutions seem to be an artifact of a theory, the attempts to physically interpret them is nevertheless meaningful. Pav$\check{s}$i$\check{c}$ was the first who observed that the RT equations can be rewritten as the Einstein equations with some modifications. To do this, one should use the fact that the equations (\ref{RT}), as mentioned above, have the form of the conservation laws for some current:
\begin{gather}
      \partial_{\mu} j^\mu_a =      0,\label{div}\\ j^\mu_a= \sqrt{-g} (G^{\mu\nu} -\varkappa T^{\mu\nu} ) \partial_{\nu} y_a \label{curr}.
\end{gather}
One can rewrite (\ref{curr}) as follows:
\begin{align}\label{RT_Ein}
      G_{\mu\nu} -\varkappa (T_{\mu\nu}+\tau_{\mu\nu})=0,
\end{align}
where
\begin{align}
      \tau_{\mu\nu} = \frac{j_\mu^a \partial_{\nu} y_a}{\varkappa\sqrt{-g}}.
\end{align}
In this way, the RT equations (\ref{RT}) are equivalent to the Einstein equations (\ref{RT_Ein}) with additional matter, whose  energy-momentum tensor  obeys the equation
\begin{gather}
      \partial_{\mu}(\sqrt{-g}\, \tau^{\mu\nu}\partial_{\nu} y^a )=0\label{RT_curr}.
\end{gather}
The embedding theory thus can be viewed as an extended General Relativity, which can be used in order to  explain the observed non-Einsteinian dynamics, usually related with the dark matter or the dark energy.

For the Friedmann case the analysis of the RT equations was performed by Davidson et al. In \cite{davids97} a model was constructed in which the extra solutions mimic the cosmological constant; in \cite{davids01} -- a model with extra solutions which emerge in a soliton manner between the early and the late universe with almost Einsteinian dynamics. The behavior of the cosmological extra solutions was also studied with the assumption of the inflation existence \cite{statja26}. It was shown that in the Friedmann approximation the "extra solutions" are strongly suppressed after the end of the inflation, if in the beginning of the inflation the initial conditions were not fine-tuned. The cosmological models were also studied within the framework of the Hamiltonian methods for systems with higher derivatives \cite{rojas09}.

In connection with these physical interpretations of extra solutions it is interesting to compare the embedding theory with the recently proposed model of "mimetic dark matter" \cite{mukhanov,Golovnev201439}. Unlike the embedding theory, this model was originally formulated as an extension of the General Relativity, but the mechanism of the appearance of extra solutions is exactly the same -- they appear as a result of the change of variables which contain differentiation. This change is analogous to (\ref{met}):
\begin{align}
      g_{\mu\nu} = \widetilde{g}_{\mu\nu} \widetilde{g}^{\alpha\beta} \partial_\alpha \phi \partial_\beta \phi, \label{mim}
\end{align}
where $\widetilde{g}_{\mu\nu}$ is a new metric, $\phi$ is a scalar field which is connected with the old metric due to (\ref{mim}) by the condition
\begin{align}
      {g}^{\alpha\beta} \partial_\alpha \phi \partial_\beta \phi = 1.
\end{align}
After varying of the Einstein-Hilbert action, in which the substitution (\ref{mim}) was made, with respect to the new variables $\widetilde{g}_{\mu\nu}$ and $\phi$, one can obtain the following field equations:
\begin{gather}
G_{\mu\nu}-\varkappa (T_{\mu\nu}+\tau \partial_\mu \phi \partial_\nu \phi)=0,\\
\partial_{\mu}(\sqrt{-g}\, \tau g^{\mu\nu}\partial_{\nu} \phi )=0,
\end{gather}
that look very similar to the RT equations in the form (\ref{RT_Ein}),(\ref{RT_curr}).

As the ordinary Einstein equations, the RT equations can be solved much more easily in the presence of a high enough symmetry group (enough for the reduction of PDEs to ODEs). The static spherically symmetric solutions together with the Friedmann models are of greatest physical interest. The RT equations were mainly analyzed  assuming  such symmetries existing.

The analysis of the static spherically symmetric case was performed by Estabrook \cite{estabrook2009}. Using the tetrad formalism for the embedding theory constructed in \cite{estabrook1999} he studied the behavior of the metric components and presented (in an implicit form) several non-Einsteinian metrics,  whose embedding function  satisfies the RT equations. However, it should be noted that, firstly, only one embedding type was considered (namely, the parabolic one), whereas it is now known that there are 6 types of embeddings with such symmetry (see below). Secondly, the non-Einsteinian metrics obtained in \cite{estabrook2009} are not asymptotically flat, what  seems quite unphysical, if one supposes a  rapid decrease  of the source mass density at the spatial infinity. Therefore it might  be interesting to perform a  more detailed and systematic analysis of the static spherically symmetric case. The next section is devoted to this case.

\section{The absence of the extra solutions\\ for 6-dimensional symmetric embeddings}
The main idea of this section can be briefly formulated as  follows. If the symmetry of a considered embedding is high enough to reduce  the RT equations to ODE, then the extra solutions are governed by a set of constants (as in the toy model (\ref{osc1})-(\ref{osc3}) discussed above). These constants vanish if we  require  the Einstein equation to be satisfied  at the initial moment (if the variable in ODE is a time) or at  some boundary (if it is  a spatial coordinate), and we can try to take the spatial infinity as this boundary. Namely, if one supposes that at the spatial infinity the metric tends to a flat one, then in that region  $G^{\mu\nu}$ vanished, so the vacuum Einstein equations are satisfied, and in the presence of a  high enough symmetry one can prove their satisfaction everywhere, i.e. the absence of extra solutions.

We will make such a proof for a  particular physically interesting case. Let us consider the existence of static spherically symmetric metrics, the embedding function for which satisfies the vacuum RT equations, but not the Einstein equations, and defines $SO(3)\otimes T^1$-symmetric 4-dimensional surface in the flat 6-dimensional ambient spacetime. We restrict ourselves to the 6-dimensional ambient spacetime due to the fact that any 4-dimensional spacetime with $SO(3)$ symmetry can be locally isometrically embedded in the spacetime of this dimension \cite{schmutzer}. The embeddings of such spacetimes to a higher-dimensional spacetime are nevertheless possible, but the behavior of the extra solutions in this case requires additional study which is beyond the scope of this paper.

The considered metrics can be written in the form
\begin{align}
      ds^2 = (1-P(r)) dt^2 - (1-Q(r))dr^2 - r^2 d\Omega^2, \label{metr}
\end{align}
where the functions $P(r)$, $Q(r)$ tend to zero at large $r$:
\begin{gather}
\lim \limits_{r\to\infty}     P(r)=0, \qquad \lim \limits_{r\to\infty}  Q(r)=0\label{PQ}.
\end{gather}
Let us suppose that there are no event horisons in the region of our interest, so in this region $P(r),Q(r)<1$.

The embeddings of the metric (\ref{metr}), which have the $SO(3)\otimes T^1$ symmetry and which are smooth in the whole considered region, can be constructed using the method described in \cite{statja27,statja29}. The embeddings of the metric satisfying the condition (\ref{PQ}) belong to the one out of the six following types.
\begin{enumerate}
      \item Elliptic:
\begin{align}
      y^0&=h(r),\nonumber \\
      y^1&=f(r)\sin(\alpha t+w(r)),\label{vlo1}\\
      y^2&=f(r)\cos(\alpha t+w(r)).\nonumber
\end{align}
\item Hyperbolic:
\begin{align}
      y^0&=h(r),\nonumber \\
      y^1&=f(r)\sinh(\alpha t+w(r)),\label{vlo2}\\
      y^2&=f(r)\cosh(\alpha t+w(r)).\nonumber
\end{align}
\item Spiral:
\begin{align}
      y^0&=k t+h(r),\nonumber \\
      y^1&=f(r)\sin(\alpha t+w(r)),\label{vlo3}\\
      y^2&=f(r)\cos(\alpha t+w(r)).\nonumber
\end{align}
\item Exponential:
\begin{align}
      y^0&=kt + h(r),\nonumber\\
      y^1&=f(r)\sinh(\alpha t+w(r)),\label{vlo4}\\
      y^2&=f(r)\cosh(\alpha t+w(r)).\nonumber
\end{align}
\item Parabolic:
\begin{align}
      y^+&=\frac{\alpha^2}{2}h(r)t^2+\alpha w(r)t+f(r),\nonumber\\
      y^1&=w(r)+ \alpha h(r) t,\label{vlo5}\\
      y^-&=h(r)\nonumber
\end{align}
(here and hereafter $y^\pm=(y^0\pm y^2)/\sqrt{2}$).
\item Cubic:
\begin{align}
y^+&=\frac{\alpha^2}{6}\left(t+h(r)\right)^3+\alpha w(r)t+f(r),\nonumber\\
y^1&=\frac{\alpha}{2}\left(t+h(r)\right)^2+w(r),\label{vlo6}\\
y^-&=t+h(r).\nonumber
\end{align}
\end{enumerate}
The three remaining components of the embedding function are the same for all the types:
\begin{align}
      y^3=r\,\cos\theta,\quad y^4=r\,\sin\theta\,\cos\phi, \quad y^5=r\,\sin\theta\,\sin\phi.
\end{align}
Here the functions $f(r), h(r), w(r)$ are related to $P(r)$ and $Q(r)$ by the equation (\ref{met}); $\alpha$ and $k$ are constants. The signature of the ambient space must be chosen for each embedding type  in such a manner that the symmetry of the embedding is preserved (e.g. for spiral embedding the signature signs of the directions $y^{1,2}$ must be the same, whereas for exponential they are  opposite) and allows it to cover the whole region outside of the possible horizon. The explicit forms of the Schwarzchild, Reissner-Nordstr$\ddot{\text{o}}$m and Schwarzschild-de Sitter metric embeddings correspond to the types listed above and can be found in \cite{statja27}, \cite{statja30} and \cite{statja32}, respectively.

The Einstein tensor constructed from the diagonal metric (\ref{metr}) is also diagonal. One can consider only two of its components $G^{00}$ and $G^{11}$ as independent, because the remaining components vanish or can be expressed through $G^{00}$ and $G^{11}$ using the Bianchi identities. For our purpose we need the explicit form of $G^{11}$ only:
\begin{gather}
      G^{11} = \frac{rP'-Q(1-P)}{(1-P)(1-Q)^2 r^2}. \label{4.5}
\end{gather}

Let us transform the left hand side of (\ref{RT}) at $a=0,1,2$ taking into account that $y^{0,1,2}$ in all considered embeddings are independent of $\theta,\phi$ and that $G^{\mu\nu}$ is diagonal and does not  depend of time:
\begin{align}\label{sp1}
 \partial_{\mu}(\sqrt{-g}\, G^{\mu\nu}\partial_{\nu}y^{a} ) =
 \sqrt{-g}\, G^{00}\partial_{0}^2y^{a}+\partial_{1}(\sqrt{-g}\, G^{11}\partial_{1}y^{a} ) =0.
\end{align}
First of all, one can observe that for all considered embeddings in any point at least one of the components $y^{0,1,2}$ for which $\partial_{0}^2y^{a}\neq 0$ exists. Therefore if $G^{11}=0$ everywhere then due to (\ref{sp1}) the value $G^{00}$ also vanishes ( assuming  that $g\ne0$). And due to the Bianchi identities all remaining $G^{\mu\nu}$ components are also equal to zero, i.e. the vacuum Einstein equations are satisfied. As a result, it is enough to prove that $G^{11}=0$ everywhere  in order to eliminate the extra solutions in the considered case.

Now we use the fact that for all embeddings there exists one of the components $y^{0,1,2}$ (we denote it $y^{*}$) such that $\partial_{0}^2y^{*}=0$. According to our definitions,  $\partial_{1} y^*=h'$. If we put $a=*$, it follows from the equation (\ref{sp1}) that
\begin{align}
\partial_{1}(\sqrt{-g}\, G^{11} h' ) =0 \quad\Rightarrow\quad
\sqrt{-g} G^{11} h'  = \tilde{C}, \label{4.499}
\end{align}
where $\tilde{C}$ is independent of $r$. Using the expression for the metric determinant
\begin{align}
g=-(1-P)(1-Q)r^4\sin^2\theta\label{4.2s}
\end{align}
and the formula (\ref{4.5}) we conclude that $\tilde{C}=C\sin\theta$, where $C=const$,
and the equation (\ref{4.499}) can be rewritten in the form
\begin{align}\label{sp2}
h'\frac{rP'-Q(1-P)}{\sqrt{1-P}\,(1-Q)^{3/2}} =C.
\end{align}

First we consider the case ${C}\neq 0$. If so, then the left hand side of (\ref{sp2}) never vanishes, and (\ref{sp2}) can be written as \begin{align}
h'=C\frac{\sqrt{1-P}\,(1-Q)^{3/2}}{P'_\xi-Q(1-P)},
\label{4.50}
\end{align}
with a new coordinate $\xi = \ln r$ such as  $rP'(r)=P'_\xi(\xi)$.
There are two possible behaviors of the function $P$ at the infinity: either it is monotonic after some value of $\xi$ (and $r$ too), in this case  considering (\ref{PQ}) we conclude that $\lim \limits_{\xi\to\infty}  P'_{\xi} =0$, or there are no such values, and after any value of $\xi$ the value of $P'_\xi$ passes through zero infinitely many times. In both cases there exists a series of unboundedly increasing values of $r$, the value of $P'_\xi$ in which tends to zero. Using this fact in (\ref{4.50}) and applying  (\ref{PQ}) again, one can conclude that $h'$ must unboundedly increase when $r\to\infty$. On the other hand, substituting each of the embedding functions (\ref{vlo1})-(\ref{vlo6}) in (\ref{met}), one can see that if (\ref{PQ}) is satisfied then  $h'$ must be at least bounded from above. Therefore the assumption $C\neq 0$ leads to the contradiction.

So we discover that $C=0$; due to the non-degeneracy of the metric this allows us to rewrite (\ref{4.499}) in the form
\begin{align}
h' G^{11}=0.
\label{sp3}
\end{align}
This means that either $h'=0$ or $G^{11}=0$. If $h'=0$ then the vacuum RT equations are simplified, and one can show that in this case the asymptotical behavior of their extra solutions is incompatible with the conditions (\ref{PQ}). As a result, we conclude that $G^{11}=0$, resulting in the satisfaction of the remaining vacuum Einstein equations.

So we have proved the absence of extra solutions of the vacuum RT equations assuming  that a $SO(3)\otimes T^1$-symmetric embedding in the 6-dimensional flat ambient space is considered and the induced metric is asymptotically flat. In other words, the only solution of the RT equations with the given restrictions is the Schwarzchild one. The exterior Schwarschild solution also remains the only solution of the RT equations outside  an arbitrary static spherically symmetric matter distribution.

\vskip 0.5em
{\bf Acknowledgments}.
The authors thank the organizers of the II Russian-Spanish Congress Particle and Nuclear Physics at all Scales and Cosmology. The work was partially supported by the Saint Petersburg State University grant N~11.38.660.2013.
The work of A.A.Sh. was also supported by the non-profit Dynasty Foundation.

%\bibliographystyle{../../IEEEtran-my}
%\bibliography{../../paston-grav-e}

\end{document}